# Metal-insulator transition in manganites: mixture of oxygen isotopes versus magnetic field.

A. Taldenkov[a*], N. Babushkina[a], A. Inyushkin[a], O. Nikolaeva[a], O. Gorbenko[b], A. Kaul[b]

[a]*Institute of Molecular Physics, RRC "Kurchatov Institute" Kurchatov sq.1, Moscow 123182, Russia*
[b]*Department of Chemistry, Moscow State University, Vorobievy Gory, Moscow 119899, Russia*

Abstract

We have investigated the effect of oxygen isotope substitution on the metal-insulator transition temperature and the resistivity of the narrow band manganite $(La_{0.25}Pr_{0.75})_{0.7}Ca_{0.3}MnO_3$ in a constant magnetic field. A set of 16 samples having different mixtures of $^{16}O$, $^{17}O$ and $^{18}O$ isotopes with average mass varying from 16.0 to 17.8 a.m.u. was studied. We have found that the magnetoresistance and the isotope effect can be linked together with a single parameter — effective magnetic field, which decreases linearly with an increase of average oxygen mass with a slope of -2 T/a.m.u. The applicability of the small polaron model is discussed.

---

[*]Corresponding author. Tel.: +7-095-196-7428; fax: +7-095-194-1994.
*E-mail address*: taldenkov@imp.kiae.ru.

**Introduction**

The observation of a giant isotope effect in manganites [1, 2] opens a new possibility to control ground state of the compounds. Manganite with narrow bandwidth $(La_{0.25}Pr_{0.75})_{0.7}Ca_{0.3}MnO_3$ (LPCMO75) is one of the well-studied compounds with a strong concurrence between various charge, magnetic and electronic degrees of freedom. As temperature decreases LPCMO75 undergoes a number of phase transformations: paramagnetic insulator - charge ordered (CO) insulator - antiferromagnetic (AFM) insulator - ferromagnetic (FM) metal or phase separated state [3]. The later can be easily changed by oxygen isotope substitution or moderate magnetic field [4]. Altering the average oxygen mass one can continuously change the FM transition temperature ($T_{FM}$) and FM volume fraction ($X_{FM}$) [5-7]. The main goal of this article is to highlight the close similarity between the effect of the external magnetic field and the effect of the oxygen mass.

**Experimental**

The isotope substituted LPCMO75 ceramic samples was investigated previously in [2-7]. The measurements were performed for 16 samples annealed either in the mixtures of two ($^{16}O$ - $^{18}O$) or three ($^{16}O$ - $^{17}O$ - $^{18}O$) oxygen isotopes. We prepared the samples with average mass varied from 16.0 to 17.8. The isotope enrichment was obtained by weighing the samples after annealing. The sample preparation and measurement procedure were described in details in [3]. Resistivity, ac-susceptibiliy, thermal conductivity and neutron diffraction experiments on these samples in zero magnetic field are presented in [5-8]. The MI transition temperature ($T_{MI}$) was determined by the maximum of logarithmic derivative $d(\ln R)/d(\ln(1/T))$.

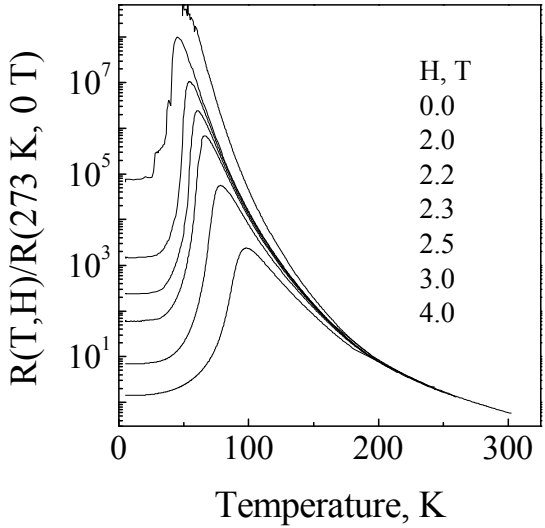

Fig.1. Temperature dependence of electrical resistivity in magnetic fields for sample with $M_O$=17.8 a.m.u..

**Results and Discussion**

The typical temperature dependencies of reduced resistivity for $LPCM^{18}O75$ ceramic at different magnetic fields are shown in Fig. 1. The data clearly demonstrate all characteristic features of manganites: the semiconducting behavior at high temperature, the field-induced FM and MI transitions and saturation of electric conductivity at low temperature. The MI transition is of percolation-type and occurs at a critical fraction of FM phase of about 20%, as was shown earlier [5, 7]. The applied magnetic field strongly increases $T_{MI}$ as does the substitution with lighter oxygen isotope.

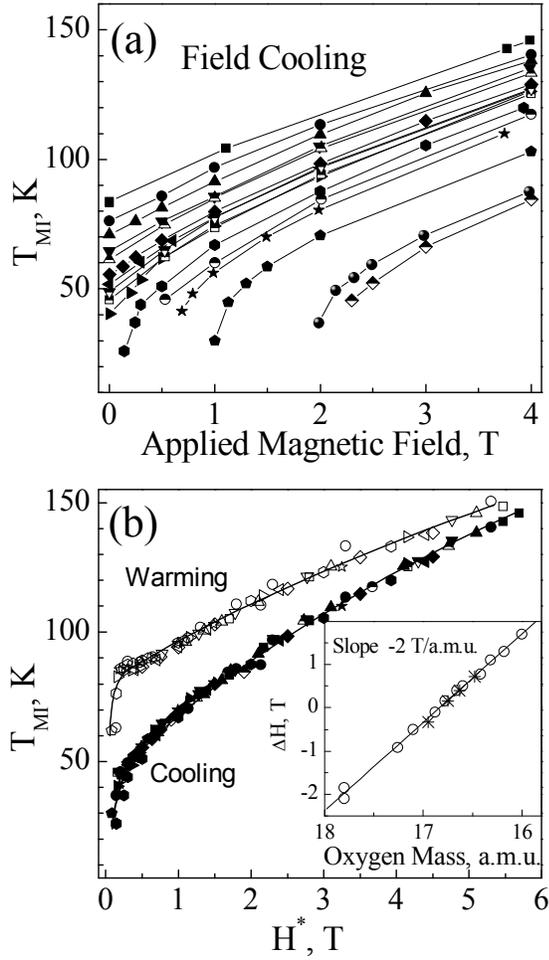

Fig.2. a) TMI in applied magnetic field for the sample with average oxygen mass: 16.0, 16.18, 16.32, 16.42, 16.48, 16.6, 16.63, 16.66, 16.75, 16.78, 16.88, 16.95, 17.1, 17.26, 17.8, 17.8 (from top to the bottom). b) Scaling behavior of TMI. The insert shows linear dependence between $\Delta H$ and average oxygen mass. Stars correspond to the samples containing $^{17}O$.

Magnetic field dependencies of $T_{MI}(H)$ for all isotopic compositions are shown in Fig. 2a. The curves corresponding to the intermediate isotope content are systematically in-between those for $M_O=16$ a.m.u. and $M_O=17.8$ a.m.u. The similarity of $T_{MI}(H)$ curves obtained for different isotope compositions suggests that all these curves can be superposed by shifting with $\Delta H$ on the x scale. It is assumed that the effect of isotope substitution is equivalent to the application of the additional magnetic field $\Delta H$, which is, in turn, proportional to the average oxygen mass. Following this approach, we redrew $T_{MI}(H)$ curves as a function of 'effective' field $H^* = H + \Delta H$ so as they all lay on the same transition line $T_{MI}(H^*)$ (Fig. 2b). It appears that parameter $\Delta H$ depends linearly on oxygen mass with a slope of about -2 T/a.m.u. (insert in Fig. 2b), being sensitive to the average oxygen mass only, but not to the isotope composition of the mixture. A large thermal hysteresis in $T_{MI}$ is clearly seen for cooling down and warming up curves. The scaling procedure with the same value of $\Delta H$ is applicable to both curves (Fig. 2b).

The low temperature residual resistivity at temperature far below $T_{MI}$ correlates strongly with fraction of FM phase in AFM matrix (Fig. 1). The value of resistance residual ratio $\rho =R(5\ K)/R(273\ K)$ for different magnetic fields and oxygen masses is shown in Fig. 3a. Despite the very strong variation of $\rho(M_O,H)$ in the samples with different isotope composition and in various magnetic fields, it can be scaled by the same manner as $T_{MI}(H^*)$ (Fig. 3b). The scaling parameter $\Delta H$ demonstrates linear mass dependence with nearly the same slope -2 T/a.m.u. (insert in Fig. 3b). The universal $\rho(H^*)$ curve manifests in implicit form the dependence of resistivity on FM fraction, which is modified by field and/or isotope. It may be well-approximated with polynomial dependence $\rho = r_0 + r_2(H^*)^2 + r_4(H^*)^4$, where $r_0$ is the resistance of the samples with maximal (saturated) amount of FM phase, $r_4(H^*)^4$ is the term characterizing resistance of 'bad' metal just above percolation threshold, and $r_2(H^*)^2$ is supposed to describe an intermediate region. The term with the fourth power $r_4(H^*)^4$ can be explained as the product of nearly quadratic dependence of resistivity versus FM fraction [5] and quadratic dependence of FM fraction versus oxygen mass [7].

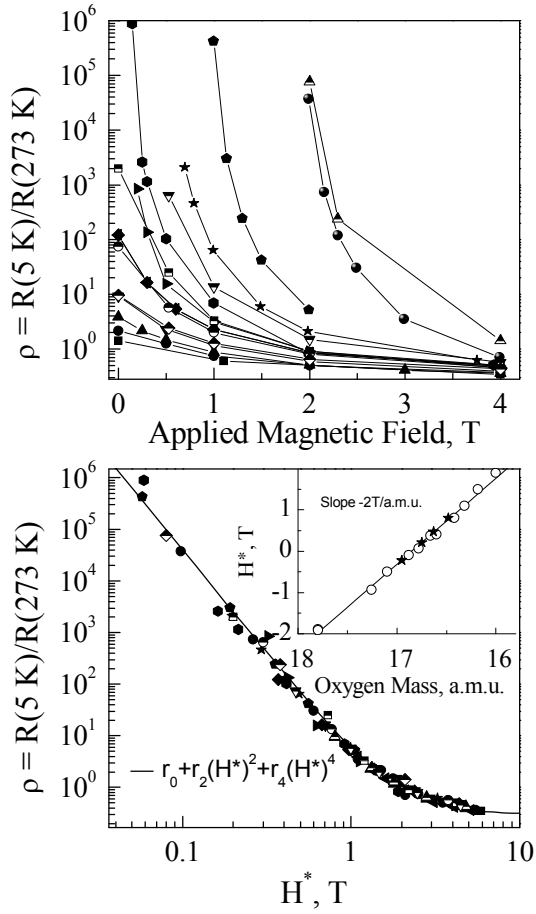

Fig.3. a) $\rho$ in applied magnetic field for the sample with average oxygen mass (Symbols are the same as in Fig. 2.).
b) Scaling behavior of $\rho$. The insert shows linear dependence between scaling parameter H* and average oxygen mass. Stars correspond to the samples containing $^{17}O$.

The temperature of MI transition in magnetic field is determined by $T_{FM}$ and $X_{FM}$, whereas $\rho$ depends on $X_{FM}$ only. So, if both $T_{MI}$ and $\rho$ scale independently with the same parameter, one may conclude that $T_{FM}$ can also be scaled with the same effective $\Delta H$.

Several authors attempted quantitatively to describe large isotope effect in manganites. The most general approach is to employ a small polaron model. It is supposed that isotope effect occurs due to the decrease of $e_g$ electronic bandwidth W. In the double exchange (DE) model $T_{FM}$ depends on effective bandwidth which is reduced via polaron narrowing factor:

$$T_{FM} \sim W exp(-\gamma E_P/\hbar\omega) \qquad (1)$$

where $\gamma$ is electron-phonon coupling constant, $\omega \sim M_O^{-1/2}$ is characteristic phonon frequency, $E_P$ is a polaron binding energy usually supposed to be independent on $\omega$. The influence on magnetic field can also be accounted in the frame of standard DE term $W \sim \langle \cos(\Theta/2) \rangle$, where $\Theta$ is an angle between Mn spins. Zhao et al. [9] discovered an important empirical constraint between isotope exponent $\alpha_0 = (\Delta T_{FM}/T_{FM})/(\Delta M_O/M_O)$ and $T_{FM}$. They showed that in a wide temperature range $\alpha_0$ obeys the law - $\alpha_0 \sim \exp(-T_{FM}/T_0)$, with parameter $T_0 = 62.5$ K, which is close to minimal observed $T_{FM}$ (Fig. 4). In our extreme case, when $\Delta T_{FM}$ is of the same order as $T_{FM}$ itself, we calculate isotope exponent as logarithmic mass derivative of experimental dependence $T_{FM}(M_O) = 890.5 - 50 M_O$ obtained in [7] from ZFC curves. It is clear (Fig.4) that our data are similar to those of [9] even for very high value of $\alpha_0$. Therefore, it might be assumed that isotopic shift has the same origin in the compounds with very different $T_{FM}$ and $\alpha_0$. Note, that the experimental relation shown in Fig. 4 is in a qualitative, but not in a quantitative agreement with eq.1 [9, 10].

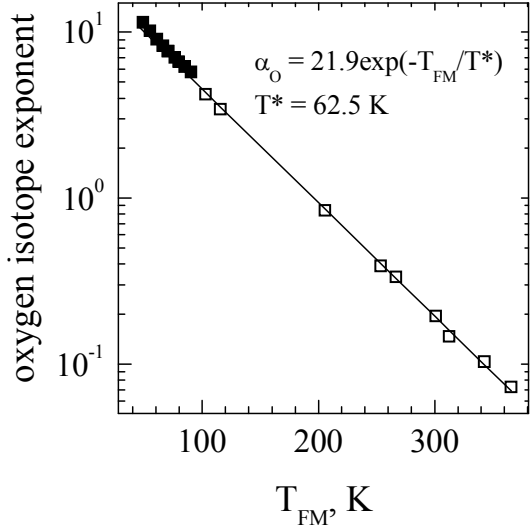

Fig.4. Isotope exponent versus $T_{FM}$.
Open symbol - data accepted from [9],
closed symbol - our data [7].

Lorenz et al. [11] tried to prove polaronic concept by varying W in optimally doped manganites $La_{0.65}Ca_{0.35}Mn^{(16-18)}O_3$ with the pressure up to 1.7Gpa. They found the linear relation $E_a = 0.5 E_P - A T_{FM}$ between the conductivity activation energy $E_a$ and $T_{FM}$. This kind of linking between $T_{FM}$, W and $E_p$ turn out to be incompatible with our results, because activation energy $E_a = 130 mev \pm 5 mev$ in our samples at $T > T_{CO}$ is nearly independent of $M_O$. Recently Plakida [12] proposed a model, which takes into account the proximity of the ferromagnetic transition to the antiferromagnetic phase with $T_{FM} \sim \{W \exp(-\gamma E_P/\hbar\omega - t_c)\}$, where $t_c$ is proportional to AFM exchange constant. To explain the experimental linear dependence of $T_{FM}(M_O)$ it is sufficient to include a weak electron–phonon interaction for free charge carriers. All models based on the change of the bandwidth also predict the isotope shift of $T_{CO}$ transition, because $T_{CO} \sim 1/W$. It is inconsistent with the data published in [7, 8], where it was stated out that $T_{CO}$ and $T_{AFM}$ are independent on oxygen isotope substitution in our samples.

As a results we conclude that the detailed quantitative description of the isotope effect is far from being settled. Recently Edwards [10] proposed that the primary isotope narrowing in W due to change in Mn-O-Mn bond angle can be strongly enhanced by increase of $E_P$ also due to modified screening of electron-phonon coupling. Thus, the scaling parameter $E_P/W$ may link CMR, isotope and pressure effects. Probably, this approach based on Holstein DE model will clear the way for better understanding of the problem.

**Acknowledgements**

This work was supported by the Russian Foundation for Basic Research (projects 02-03-33258, 03-02-16954, and 04-02-16991).

**References**

[1]  M. Zhao, K. Conder, H. Keller, and K.A. Müller, Nature (London) 381 676 (1996).
[2] N.A. Babushkina, L.M. Belova, O.Yu. Gorbenko, A.R. Kaul, A.A. Bosak, V.I. Ozhogin, K.I. Kugel, Nature (London) 391 (1998) 159.
[3] A.M. Balagurov, V.Yu. Pomjakushin, D.V. Sheptyakov, V.L. Aksenov, N.A. Babushkina, L.M. Belova, O.Yu. Gorbenko, A.R. Kaul, Eur. Phys. J. B 19 (2001) 215.
[4] N.A. Babushkina, L.M. Belova, V. I. Ozhogin, O.Yu. Gorbenko, A.R. Kaul, A.A. Bosak, K.I. Kugel, D.I. Khomskii, J.Appl. Phys. 83 (1998) 7369.
[5] N.A. Babushkina, A.N. Taldenkov, L.M. Belova, E.A. Chistotina, O.Yu. Gorbenko, A.R. Kaul, K.I. Kugel, D.I. Khomskii, Phys.Rev.B 62 (2000) R6081.
[6]  N.A. Babushkina, A.N. Taldenkov, E.A. Chistotina, A.V. Inyushkin, O.Yu. Gorbenko, A.R. Kaul, K.I. Kugel, D.I. Khomskii, J.Magn. Magn.Mater. 242–245 (2002) 640.
[7] A.N. Taldenkov, N.A. Babushkina, A.V. Inyushkin, J.Magn. Magn.Mater. 258–259 (2003) 271.
[8]  A.M. Balagurov, V.Yu. Pomjakushin, D.V. Sheptyakov, N.A. Babushkina, Appl. Phys. A 74 [Suppl.] (2002) S1737.
[9]   M. Zhao, K. Conder, H. Keller, and K.A. Müller, Phys. Rev. B 60 (1999) 11914.
[10] D.M. Edwards, Adv. Phys. 51 (2002) 1259.
[11]  B. Lorenz, A.K. Helman, Y. S. Wang, Y.Y. Yue, and C.W. Chu, Phys. Rev. B63 (2001) 144405.
[12]  N.M. Plakida, JETP Letters 71 (2000) 493.